\RequirePackage{fix-cm,amsmath}
\documentclass[twocolumn]{svjour3}          
\smartqed  
\usepackage{graphicx}
\usepackage{newtxtext,newtxmath}     
\usepackage[utf8]{inputenc}
\usepackage[nolist]{acronym}
\usepackage{eurosym}
\PassOptionsToPackage{hyphens}{url}\usepackage{hyperref}
\usepackage{amsmath}
\usepackage{xcolor}
\usepackage{tikz}
\usepackage{cite}
\usepackage{booktabs}
\usepackage{orcidlink}
\usepackage{float}
\usepackage{longtable}
\usepackage{float}

\usepackage{url}

\usepackage[utf8]{inputenc}
\usepackage{graphicx}
\usepackage{enumitem}

\usepackage[T1]{fontenc}

\journalname{\bfseries{Informatik Spektrum}}
\journalinfo{It can be accessed at \href{https://link.springer.com/article/10.1007\%2Fs00287-020-01321-z\#Ack1}{https://link.springer.com/article/10.1007\%2Fs00287-020-01321-z\#Ack1}}

\begin{document}
\sloppy

\newcommand{\change}[1]{\textcolor{red}{#1}}

\begin{acronym}
\acro{pow}[PoW]{Proof-of-Work}
\acro{pos}[PoS]{Proof-of-Stake}
\acro{poa}[PoA]{Proof-of-Authority}
\end{acronym}

\title{Recent Developments in Blockchain Technology and their Impact on Energy Consumption}

\author{Johannes Sedlmeir \and
        Hans Ulrich Buhl \and
        Gilbert Fridgen \and
        Robert Keller }

\authorrunning{J. Sedlmeir et al.} 

\institute{Johannes Sedlmeir\,$^{\orcidlink{0000-0003-2631-8749}}$ (Corresponding author) \at   
             FIM Research Center, University of Bayreuth\\
             Project Group Business \& Information Systems Engineering of the Fraunhofer FIT \\
             Bayreuth, Germany \\
             \email{johannes.sedlmeir@fit.fraunhofer.de}         
          \and
          Hans Ulrich Buhl \at
              FIM Research Center, University of Augsburg\\
              Project Group Business \& Information Systems Engineering of the Fraunhofer FIT \\
              Augsburg, Germany \\ 
              \email{hans-ulrich.buhl@fim-rc.de}
          \and
          Gilbert Fridgen\,$^{\orcidlink{0000-0001-7037-4807}}$ \at
              SnT - Interdisciplinary Center for Security, Reliability and Trust \\
              University of Luxembourg \\
              \email{gilbert.fridgen@uni.lu}   
          \and
          Robert Keller\,$^{\orcidlink{0000-0001-7097-1724}}$ \at
              FIM Research Center, University of Augsburg\\
              Project Group Business \& Information Systems Engineering of the Fraunhofer FIT \\
              Augsburg, Germany \\
              \email{robert.keller@fim-rc.de}
}

\maketitle
\begin{abstract}
The enormous power consumption of Bitcoin has led to undifferentiated discussions in science and practice about the sustainability of blockchain and distributed ledger technology in general. However, blockchain technology is far from homogeneous -- not only with regard to its applications, which now go far beyond cryptocurrencies and have reached businesses and the public sector, but also with regard to its technical characteristics and, in particular, its power consumption. This paper summarizes the status quo of the power consumption of various implementations of blockchain technology, with special emphasis on the recent ``Bitcoin Halving''~and so-called ``zk-rollups''. We argue that although Bitcoin and other proof-of-work blockchains do indeed consume a lot of power, alternative blockchain solutions with significantly lower power consumption are already available today, and new promising concepts are being tested that could further reduce in particular the power consumption of large blockchain networks in the near future. From this we conclude that although the criticism of Bitcoin's power consumption is legitimate, it should not be used to derive an energy problem of blockchain technology in general. In many cases in which processes can be digitised or improved with the help of more energy-efficient blockchain variants, one can even expect net energy savings.

\end{abstract}
\section*{Introduction}
\label{sec:introduction}
In leading German print media one can find statements that the Bitcoin system consumes about as much electricity as the Federal Republic of Germany, tendency rising~(Frankfurter Allgemeine Zeitung, 2020-06-06~\cite{faz2020bitcoin}).
On the other hand, an article was published in the magazine \emph{Nature Climate Change} in 2018, according to which if Bitcoin is adopted on a large scale, the emissions caused by it alone could lead to global warming of more than 2$^\circ$C in the next three decades~\cite{Mora:2018:Bitcoin}. The FAZ article was modified on our initiative shortly after its publication in the online version, and the Nature article was followed by a controversial scientific discussion about the sense of the underlying assumptions. Nevertheless, such publications lead to an incorrect impression in the public regarding the ecological consequences of Bitcoin and to an even more problematic generalisation to blockchains.

In essence, the statement that Bitcoin and also many other cryptocurrencies cause an enormous power consumption is correct and important and has been analysed in detail in numerous publications, including the journals \emph{Joule}  \cite{deVries:2018:Bitcoin, stoll2019carbon,gallersdorfer2020beyond} and \emph{Nature Sustainability}~\cite{Krause:2018:Quantification}.
Frequently, however, it is precisely these striking statements that remain present, are taken out of context, are incorrectly generalised or used for lines of argumentation that testify to a lack of understanding of the fundamental interrelation between the high electricity consumption of some cryptocurrencies and economic as well as technical parameters. For example, Bitcoin's electricity consumption does not necessarily increase steadily nor does it grow significantly with the number of transactions processed per time unit. Moreover, blockchain technology is mentioned in the same breath as Bitcoin so frequently, both in public reporting and, to some extent, in the scientific community, that certain prejudices regarding the power consumption of blockchain technology have become generally established.

In fact, there are now numerous cryptocurrencies based on technically significantly modified blockchain variants with completely different characteristics with regard to their power consumption. The situation is similar for a large number of implementations of blockchain-based platforms for cross-organisational processes in business and the public sector. In Germany, for example, there are projects by automobile manufacturers in the supply chain~\cite{miehle2019partchain} or the Federal Office for Migration and Refugees~\cite{Rieger:2019:BAMF}. As the topic of sustainability is rightly very present in politics and economy~\cite{Gimpel2019Sustainability}, the question of electricity consumption and the sustainability of blockchain technology in general is very often asked in the context of blockchain-related projects for the reasons described above. The presence of this electricity consumption stigma could therefore significantly impede the adoption of blockchain technology and, thus, the exploitation of its benefits~\cite{Beck:2018:Governance}.

Accordingly, in this paper we want to give a comprehensive overview of the electricity consumption of blockchain technology in general in order to provide a solid basis for the general discourse. To this end, we first describe well-known estimates for the energy consumption of Bitcoin, but expand on these estimates by a detailed discussion of the recent Bitcoin Halving, which reveals many of the fundamental interrelations. In the  Bitcoin Halving event, which periodically takes place approximately every four years, the number of new Bitcoins created per block and serving as a reward for the miner is halved. This ensures that the number of existing Bitcoins remains limited (geometric series). The aim of this construction is to reduce inflation.
On the other hand, we are also investigating a larger part of the very heterogeneous spectrum of blockchains than just some cryptocurrencies that are technically closely related to Bitcoin. With this we extend an article~\cite{sedlmeirBISE} published by us in the magazine \emph{Business \& Information Systems Engineering} on the energy consumption of blockchains, which already discusses some of the issues addressed in this article and focuses more on the sustainability discussion of blockchain technology applications beyond cryptocurrencies. In comparison, we will go into more detail on some aspects only briefly discussed there. In particular, we will quantitatively analyse the implications of using so-called zk-rollups on the power consumption of blockchains in addition to Bitcoin Halving.

Despite the fact that blockchain technology is used in a much wider range of applications than in Bitcoin and cryptocurrencies, Bitcoin also plays a central role in this article. This is due to its problematic high energy consumption. We believe, however, that other applications of blockchain technology are much more important in the long run.

\section*{Basics of Bitcoin and Blockchain Technology}
The Bitcoin blockchain was developed to create a decentralised electronic currency system. The transfer of assets -- in contrast to the transfer of information -- is not readily possible bilaterally in electronic form, since electronic objects can be copied practically as often as desired without effort. Therefore, although the information contained in the electronic objects may be valuable in itself, no value is transferred by the transmission or storage of the electronic object itself~\cite{faulkner2019theorizing}. Accordingly, for the electronic transfer of values within a certain group, a so-called ledger that is accepted by all members and contains the ownership relationships is required. A change in such ownership in an electronic register can therefore be understood as the electronic transfer of values (``transaction'').

Traditionally, trusted third parties, e.g., banks in the monetary context, have controlled such digital ledgers in the form of databases. In contrast, the cryptocurrency Bitcoin, which was presented in a white paper in 2008~\cite{Nakamoto:2008:Bitcoin} and was subsequently implemented and went into operation in 2009, is based on the decentralised management of the corresponding ledger by redundant and synchronised (``physically decentralised, logically centralised'') maintenance of the ledger on all participating computers (``nodes''). This means that the validity and execution of transactions is now decided by all participants in the Bitcoin network rather than by a single central authority. Proving ownership of units of the cryptocurrency (token-based model) or authorising payments (account-based model) is done with the help of a public key infrastructure and corresponding digital signatures. For the purpose of majority decision making, the Bitcoin network requires a so-called consensus mechanism whereby the nodes decide which new transactions to include, and in what order. 

In principle, such replicated state machines, which guarantee the security and functionality of a distributed network even in the presence of system failures or Byzantine errors, have been intensively researched~\cite{lamport1982byzantine} since 1982 and then practically implemented with Paxos\cite{Lamport1998Part} and PBFT~\cite{castro1999practical}. Consensus could be found based on elections, following the principle of ``one node, one vote''. What is new about Bitcoin, however, is that not only predefined nodes can participate in the network and the consensus process, but anyone who wants to can do so. This is known as a public permissionless system. In it, the election-based process just described is not possible, because an attacker who wants to overrule the system would only have to register a sufficient number of accounts on the network, which would be possible for them without significant costs (this is called ``Sybil attack'', see, e.g., ~\cite{douceur2002sybil}).

A permissionless system like Bitcoin must therefore tie the weight of a vote to a scarce resource in order to prevent such attacks. With Bitcoin and many other cryptocurrencies this is done via the so-called ``\ac{pow}'', i.e., the weight of the vote is linked to proven, performed computing work and, thus, energy. \ac{pow} involves finding a random number, the so-called nonce, so that the hash value of the nonce -- together with other data -- takes on a certain form. In the case of Bitcoin, this is the requirement that the integer representation of the hash value is smaller than a certain upper limit. The choice of this upper limit thus defines a measure of complexity, the so-called difficulty, of this cryptographic puzzle. The difficulty is indirectly proportional to the probability that a randomly chosen nonce leads to a hash value of the desired form. This method of proving computation effort has been known for a long time and has been discussed, for example, in Hashcash for preventing spam~\cite{back2002hashcash}.

Participation in the \ac{pow} consensus mechanism, i.e. the search for appropriate nonces, is therefore associated with costs, so that an economic incentive must be created for participation in mining: Whoever finds a nonce that, together with a bundle of transactions, leads to a hash value of the required form may also register a reward for themselves, more precisely, a certain number of new Bitcoins created for this purpose (``block reward''). The corresponding block can then be communicated to the other participants in the blockchain network and thus be attached to the existing chain, which means that the corresponding transactions are executed. It should be mentioned that due to the competition in Bitcoin mining, participation with CPUs has long since become unprofitable, as specialised hardware, so-called ASICs, have been developed that can calculate \mbox{hashes} by orders of magnitude faster and more energy-efficiently than CPUs and GPUs~\cite{deVries:2018:Bitcoin}. The difference in Bitcoin is so significant that even the world's 500 largest supercomputers taken together can probably only achieve a small part of the current hash rate of Bitcoin, which is mainly based on ASICs -- and can only conduct mining at a considerable financial loss. 

To prevent undetected manipulation of the system, Bitcoin uses a data structure that makes it very easy to detect and locate subsequent changes, namely Merkle Trees. In order to reduce the effort involved in finding a consensus, transactions, metadata, nonce and a hash pointer to the previous block are combined in one block. The resulting append-only structure (``chain'') is given the property that changing only one single transaction either leads to the inconsistency of a single block (wrong Merkle root) or all hash pointers from the manipulated block on would have to be changed as well. Due to the requirement that the hash value of each block must have the form described above, finding such blocks and thus an alternative chain of hash pointers is very computationally intensive, so that the system is secure as long as a large part of the hash rate is provided by ``honest'' nodes (for a detailed presentation of attack scenarios see~\cite{Eyal:2014:Majority}). The data structure of blocks and hash pointers is generally characteristic of blockchain technology, which in turn is a special case of so-called distributed ledger technologies. However, blockchain technology is usually understood to comprise not only this data structure, but also the existence of a consensus mechanism that both allows agreement on the addition of new transactions and ensures that no subsequent changes can be made to the blockchain. For this purpose, however, other methods can be used instead of competing for a solution of computationally intensive puzzles. These are usually based on digital signatures which, depending on the consensus mechanism, are created by network participants either according to fixed rules or \mbox{(pseudo-)randomly}. We will briefly describe some of these variants in the discussion of alternative consensus mechanisms.
\section*{Estimates for the Energy Consumption of PoW Blockchains}
As described above, the incentive for participating in mining in the Bitcoin blockchain and generally in \ac{pow}-based blockchains is a  reward in the form of a fixed number of units of the corresponding (``native'') cryptocurrency for a node that finds the next block. Due to the strong price increases of cryptocurrencies with a peak at the end of 2017 and a market capitalisation of briefly over 300\,billion and since then always over 50\,billion US dollars of Bitcoin alone, there is and has been a strong economic incentive to participate in mining. In order to maintain the functionality (and also the security) of the \ac{pow} blockchain network, the time span in which a new block is usually found must be kept constant, i.e., the difficulty of the hash puzzle must be adjusted according to the current hash rate. This leads to a correspondingly high power consumption of \ac{pow} based cryptocurrencies.

The exact determination of the power consumption in a public permissionless \ac{pow} blockchain is in general very difficult because usually neither the computing power used in mining nor the corresponding hardware can be determined for each individual participant. However, a lower limit for the power consumption of Bitcoin and all other \ac{pow}-based blockchains can easily be determined from the indirectly observable average computing power, i.e. the hash rate, and the most energy-efficient mining hardware on the market~\cite{Vranken:2017:Sustainability, deVries:2018:Bitcoin}. In doing so, one can estimate the expected value of the hash rate from the publicly visible, current difficulty of the hash puzzle and the number of solutions communicated in the form of new blocks. In Bitcoin, by construction of the protocol, a new solution of the hash puzzle is found around every 10 minutes on average, and the probability that a random hash value meets the requirements was about 1\,:\,6$\times$10$^{\mathrm{22}}$ at the beginning of 2020. For Bitcoin, SHA-256 is used as the hashing algorithm; modern ASICs achieve hash rates in the order of $10^\mathrm{14}$ hashes per second at a power of a few thousand watts. This results in a lower limit for the electricity demand of Bitcoin in early 2020 of about 60\,TWh per year~\cite{sedlmeirBISE}, which corresponds to the annual electricity consumption of about 15~million households.

An upper limit for the electricity consumption caused by mining can also be estimated, as long as it is assumed that all participants act rationally, meaning that they aim to make a profit by participating in mining. This may not be true for all participants, but the vast majority of the computing power for Bitcoin and other relevant \ac{pow} cryptocurrencies is provided by companies or groups specialising in mining (``pools'')~\cite{romiti2019deep}, so this assumption seems reasonable for them. The value of the economic incentive, i.e. the new Bitcoins produced by mining, must be at least as high on average as the costs caused by mining, e.g. for electrical energy and hardware, and, thus, in particular higher than the expenses for electricity. A lower limit for the costs of electrical energy in countries with significant mining involvement is usually set at USD~0.05 per kilowatt-hour~\cite{ElectricityPrices, MiningPoolDistribution}, and this results in an extrapolated upper limit for electricity consumption of about 120\,TWh per year for early 2020 at a Bitcoin price of just under USD~10\,000, which corresponds to about 20\,\% of the German electricity consumption~\cite{sedlmeirBISE}.

For other well-known \ac{pow} blockchains, such as Ethereum, Bitcoin Cash, Bitcoin~SV and Litecoin (these are the largest \ac{pow}-based cryptocurrencies after Bitcoin in terms of market capitalisation), the same estimation formulas apply as for Bitcoin, except that there are other hashing algorithms, specialised mining hardware and parameters such as average block times and block rewards. In total, the electricity consumption of the four cryptocurrencies mentioned above is between 10~and~40\,TWh per year, which is significantly lower than that of Bitcoin. It is also noted that due to generally similar parameters, there is a very high correlation between market capitalisation and electricity consumption for different \ac{pow} cryptocurrencies. As the market capitalisation of Bitcoin is higher than the cumulative market capitalisation of all other cryptocurrencies, it can be assumed that the cumulative electricity consumption of all \ac{pow} cryptocurrencies is not much more than twice that of Bitcoin, and a ``Best Guess''~ is at a factor of approximately~1.5~\cite{sedlmeirBISE,gallersdorfer2020beyond}.

An important observation for \ac{pow} cryptocurrencies is that their power consumption cannot be reduced in the long term by increasing the energy efficiency of hardware: On the one hand, this can be seen from the fact that the estimation of the upper bound depends only on electricity prices and not on total computing power. The reason for this is that in the long term, all miners would switch to more energy-efficient hardware as long as mining is profitable. Accordingly, as described above, the total computing power of the network will increase until the balance of revenue and expenses sides is approximately restored.

Due to the requirement to enable as many nodes as possible to participate in cryptocurrencies and the redundant execution of all transactions, the technical requirements for participation, i.e. network bandwidth and storage space, must be kept as low as possible. Since the ``slowest''~permitted node dictates the performance of the system, Bitcoin and other cryptocurrency systems can only process a few transactions per second -- currently, the storage space required for the complete Bitcoin blockchain requires just under 300\,GB and is growing by about 60\,GB per year; a multiple of transactions per time unit would also multiply the growth accordingly. Hence, simply dividing the power consumption by the number of transactions in \ac{pow}-based cryptocurrencies yields an enormous amount of energy on a per transaction basis:  For Bitcoin, the electricity consumption for a single transaction would then amount to several hundred kWh and thus correspond to the electricity consumption of an average German household in several weeks to months. This leads to the frequent criticism of Bitcoin's sustainability described above. For other \ac{pow}-based cryptocurrencies, one gets significantly lower energy consumption per transaction, but this is still orders of magnitude more energy-intensive than, for example, a conventional booking in the banking system. It is essential to understand, however, that the number of transactions processed has no effect on the total network's mining-related power consumption, since in theory the blocks could be arbitrarily enlarged~\cite{Dittmar:2019:Bitcoin}. Thus, the metric ``energy per transaction''~for \ac{pow}-based cryptocurrencies needs to be considered quite carefully. Nevertheless, given the performance of Bitcoin and other current \ac{pow} blockchains, their power consumption can certainly be regarded disproportionate.

\section*{Outlook: Implications of the recent Bitcoin Halving}

\begin{figure}[!tbh]
     \centering
     \includegraphics[width=1.0\linewidth, trim=0cm 0cm 0cm 0.5cm, clip]{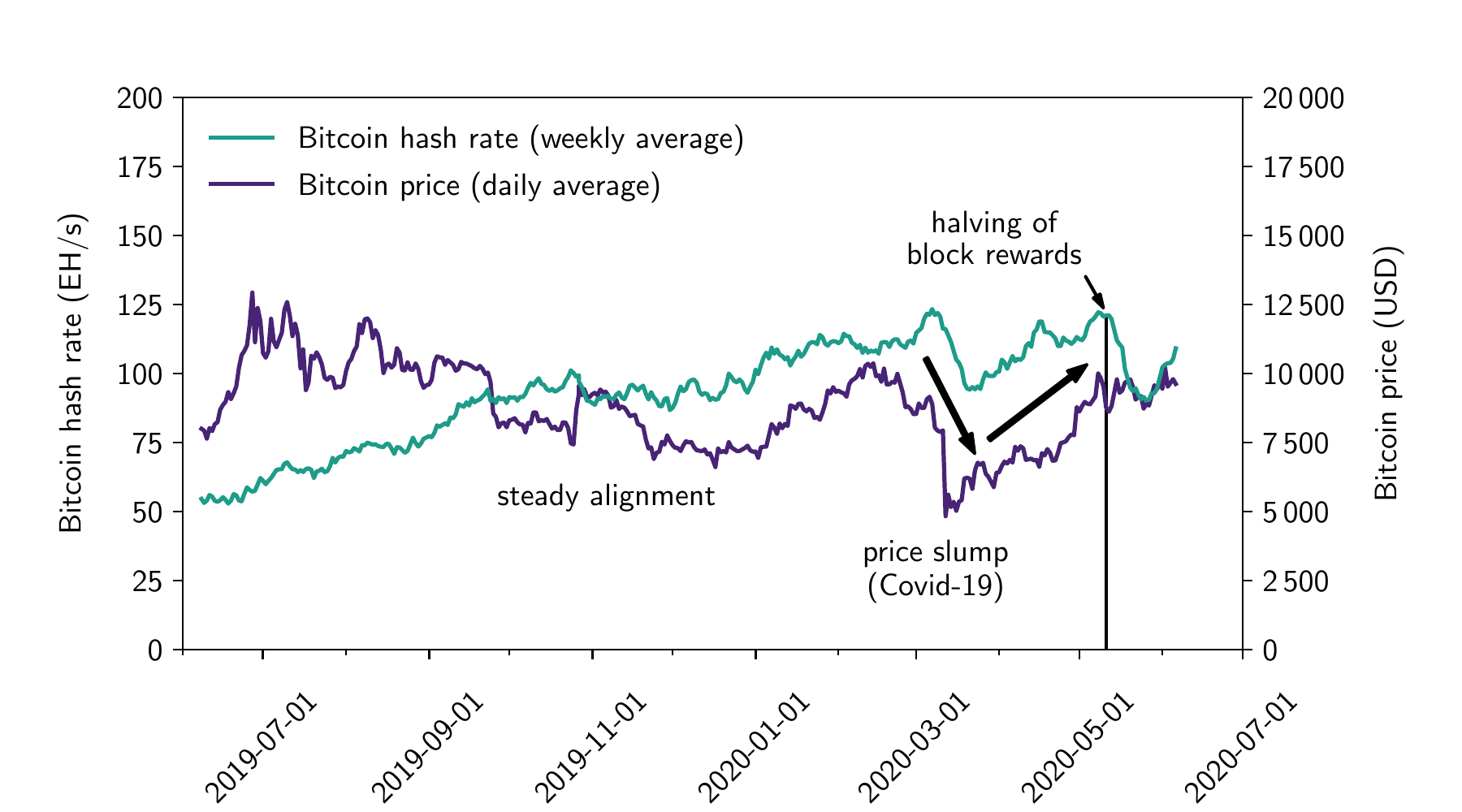}
     \caption{Hash rate and price development for Bitcoin. 1\,EH denotes 10$^\mathrm{18}$~hashes. Data retrieved from~\cite{Blockchain.com}.}
    \label{fig:hashrate}
\end{figure}

In the following, a more detailed analysis of Bitcoin's electricity consumption will be carried out by analysing the recent Bitcoin Halving and deriving implications for the long-term development of electricity consumption. The comparison of the development of Bitcoin prices and hash rate over the past 12 months, as shown in Figure~\ref{fig:hashrate}, suggests that the upper bound described above is indeed a fairly good estimate of actual electricity consumption: With relatively stable Bitcoin prices until March 2020, the observed hash rate increased continuously; apparently, the initiation or expansion of mining activities, which is associated with the procurement of appropriate hardware, was considered worthwhile. However, a fall in the price of Bitcoin in early March 2020 in the context of a generally weak stock market sentiment due to the Covid-19 pandemic was accompanied by a slightly less pronounced but nevertheless significant decrease in the hash rate. This could be explained by the fact that due to the reduction in the value of Bitcoin and thus the level of the mining incentive, miners with higher variable costs, for example due to obsolete hardware or high electricity prices, were forced out of mining here for a short period of time. Thereafter, aligned with the Bitcoin price, the hashrate also rose back to the previous level. However, the Bitcoin halving on 11 May 2020, an event that is similarly scheduled in many \ac{pow} blockchains and which takes place approximately every four years in the case of Bitcoin, caused a permanent halving of the block rewards and a corresponding reduction of the economic incentive to mine. As Bitcoin prices remained largely constant, the hash rate fell significantly, similar to the previous situation.

Surprisingly, however, just one week later, the hash rate rose again significantly, without Bitcoin prices rising to a similar extent. This could be due to the following reasons:

\begin{itemize}
    \item A closer look at the expected profits from mining must also include transaction fees received by the producer of a new block -- especially after halving, these fees made up to 20\,\% of the reward on some days, and the transaction fees have also increased after the halving.
    \item The Difficulty is not adjusted in real time, but only around every 14 days. While with the Difficulty before the halving it might not have been worthwhile to participate in mining, this might have changed after the Difficulty was adjusted for the first time after the halving and thus significantly reduced (it takes a while for the system to regain its balance).
    \item In China, the rainy season begins in May and June in some regions, so that much cheaper electricity is available through hydropower, and some mining pools were  offering electricity for just under 0.03\,USD/kWh at that time; especially since competition has become much fiercer in the face of declining mining revenues~\cite{bloomberg2020generationcosts}.
    \item More and more modern, energy-efficient hardware is being purchased and used, significantly reducing variable costs.
\end{itemize}

\begin{figure}[!tbh]
    \centering
    \includegraphics[width=1.0\linewidth, trim=0cm 0cm 0cm 0.5cm, clip]{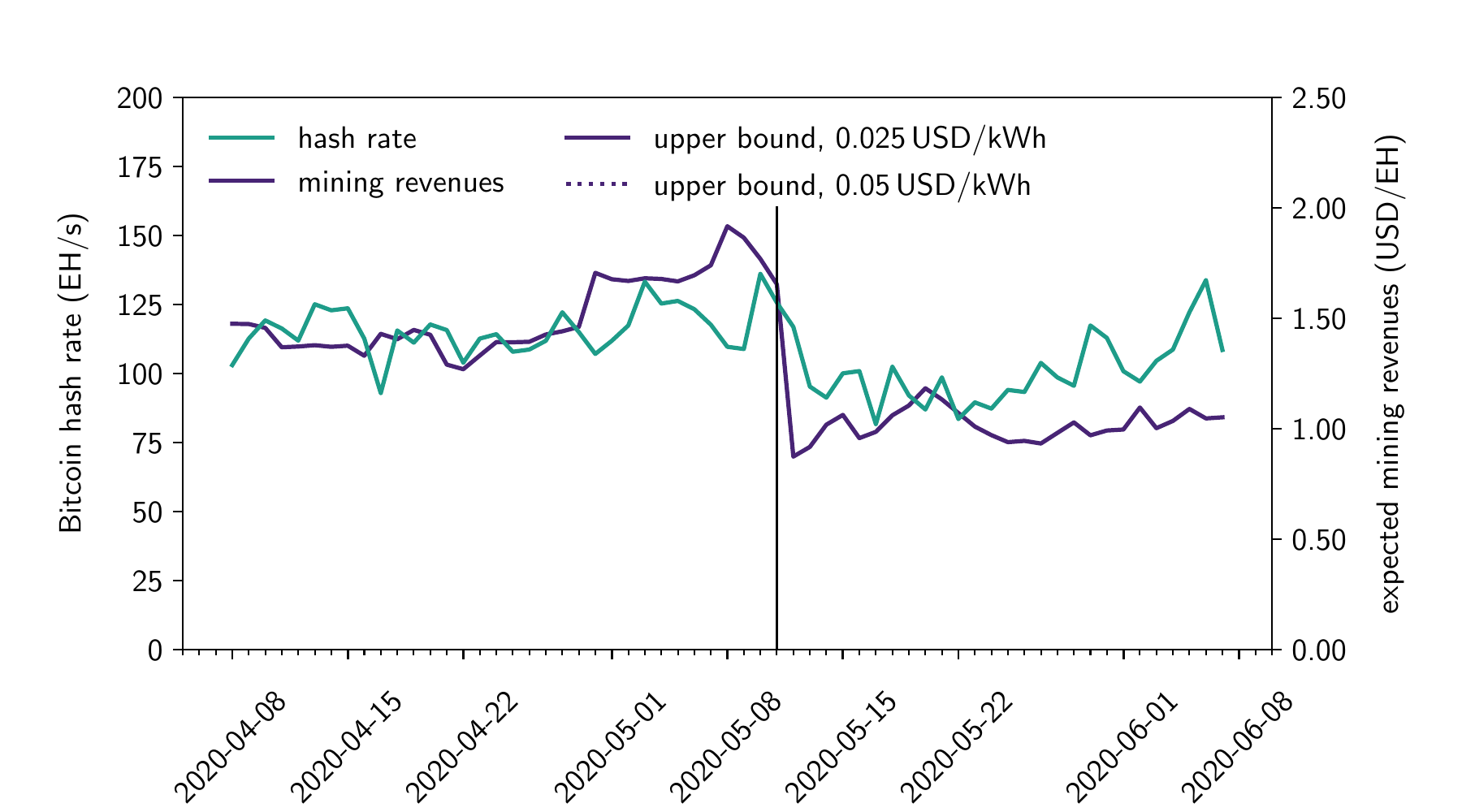}
    \caption{Hash rate and expected mining rewards for Bitcoin. Data retrieved from \cite{Blockchain.com}.}
    \label{fig:second_order}
\end{figure}

To investigate these relationships, we analyse the relationship between the economic incentive to participate in mining, in the form of an expected mining revenue in USD per 10$^\mathsf{18}$ calculated hashes, and the actual participation in the form of the hash rate, in the period around the halving with a more accurate model. This also takes into account the actual current block time or difficulty as well as the transaction fees received by the generator of a new block in addition to the block rewards. The resulting curves are shown in figure~\ref{fig:second_order}. The correlation between expected profits and the hash rate is approx.~0.57. This suggests that some miners are already deciding in real time or in the short term whether or not it is worthwhile for them to participate in mining because they have already operated close to cost neutrality before halving. Apart from the irregularities around halving, lower correlations have been observed, especially in the last two weeks analysed, which could be due to the changes in electricity tariffs in China's mining pools described above.

\begin{figure}[!tbh]
    \centering
    \includegraphics[width=1.0\linewidth]{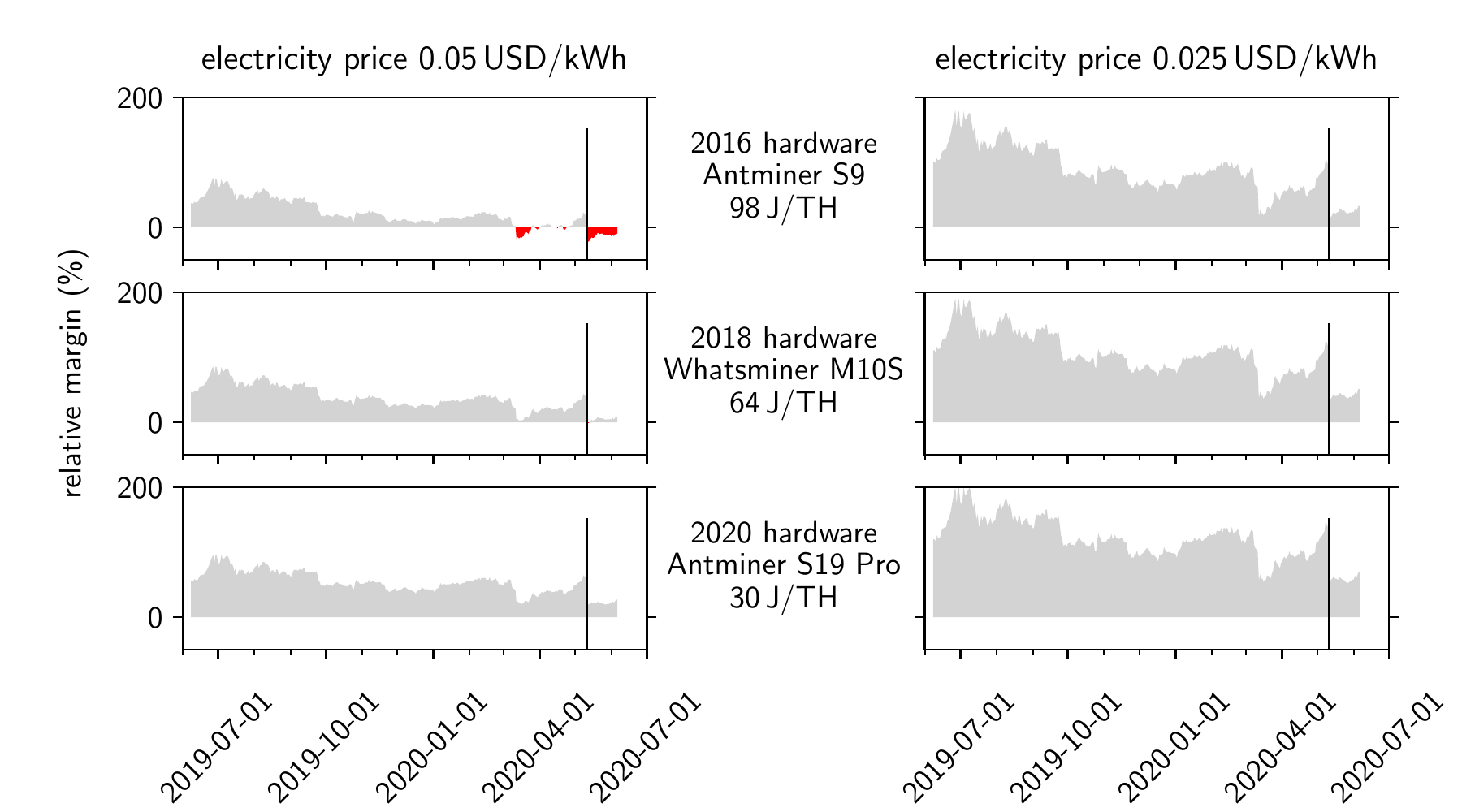}
    \caption{Profitability of mining Bitcoin. Data retrieved from\cite{Blockchain.com} and\cite{ASICProfitability}.}
    \label{fig:margin}
\end{figure}

In order to analyse the influence of electricity tariffs in more detail, we illustrate in Figure~\ref{fig:margin} the influence of mining hardware and electricity prices on the relative margin, i.e., the ratio of mining profits (i.e. the difference between mining revenues and electricity costs) to the electricity costs of mining in detail. Only electricity consumption is taken into account; other variable costs and investments (e.g. for hardware procurement) are ignored. Thus, the data shown represents an upper limit of the relative margin. The reference hardware used was the popular Bitmain Antminer S9 (11.5\,Th) for hardware newly launched in 2016, the MicroBT Whatsminer M10S for 2018, and the Bitmain Antminer S19 Pro (110\,Th) for 2020~\cite{ASICProfitability}. At the time of their market launch, these probably corresponded to the most energy-efficient mining hardware for Bitcoin. The vertical line shows the time of the Bitcoin halving, as in Figure~\ref{fig:hashrate} and Figure~\ref{fig:second_order}. In fact, at electricity prices of 0.05\,USD/kWh, the halving can force old, less energy-efficient hardware out of the market in the short term, whereas more modern, more energy-efficient hardware remaines profitable and, at lower electricity prices, mining with older hardware also makes economic sense.

\begin{figure}[!tbh]
    \centering
    \includegraphics[width=1.0\linewidth]{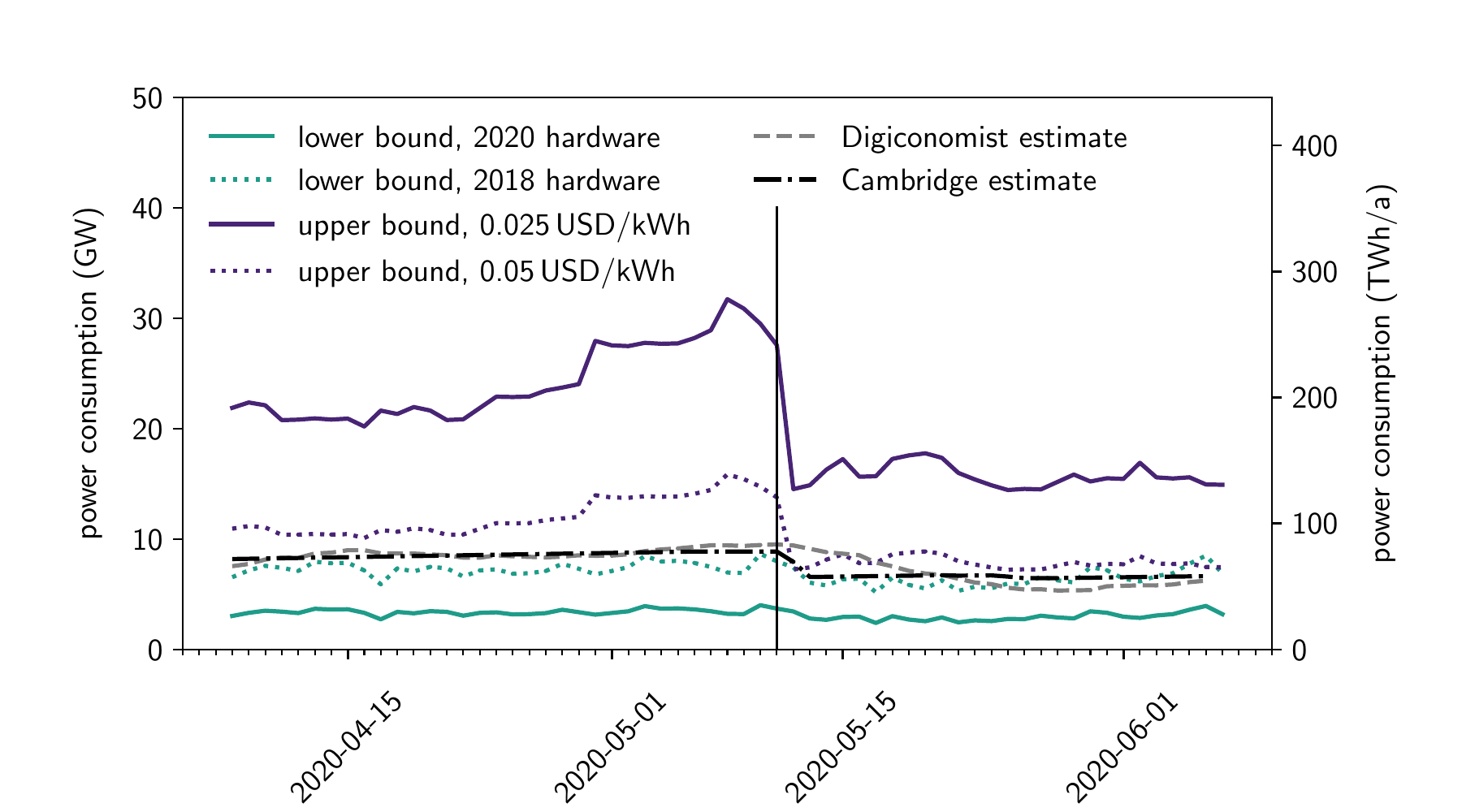}
    \caption{Estimates for the electricity consumption of Bitcoin.}
    \label{fig:energy_consumption_bitcoin}
\end{figure}

Based on the observations, it is obvious that the derived upper bound is now indeed a good estimator of the actual electricity consumption of Bitcoin. Figure~\ref{fig:energy_consumption_bitcoin} shows the electricity consumption of Bitcoin resulting from the different scenarios. The lower bound with 2020 hardware and the upper bound with 0.025\,USD/kWh can be considered very reliable, as it is unlikely that significant mining activities with more efficient hardware than the most advanced available on the market or electricity costs below 0.025\,USD/kWh will occur. More realistic, however, on the basis of actual hardware and electricity prices typical of the network, are the two dashed barriers. Although the estimates of Digiconomist~\cite{digiconomist2020bitcoin} and Cambridge~\cite{cambridge2020bitcoin} seem plausible and fit well within these realistic limits given the safe upper and lower barriers, they may not sufficiently explain the collapse of the hash rate as a result of the price collapse and the Bitcoin halving. In this respect, one could well expect that the actual hash rate before the halving was rather oriented towards the upper limit of 0.05\,USD/kWh and that after the halving, due to the increased competition from cheap electricity tariffs, mining hardware that was initially forced out of the market was also used again and thus the actual electricity consumption even rose above the upper limit of~0.05\,USD/kWh.

However, one could expect that this is only a temporary effect and that after the end of the rainy season, electricity consumption will tend to return to the upper limit of 0.05\,USD/kWh. Even if electricity prices would stay lower than~0.025\,USD/kWh, it can be assumed that in this case it would be electricity from renewable energy sources, as their marginal cost could also be zero, whereas the cost of electricity generation from fossil or nuclear fuels is unlikely to fall below 0.025\,USD/kWh. In this respect, even exceeding the ``safe'' upper barrier due to locally or temporarily low electricity prices would probably not result in a higher CO$_2$ footprint for Bitcoin than with the upper barrier at 0.025\,USD/kWh.

Assuming that the economic conditions remain unchanged, i.e., the prices for Bitcoin and electricity as well as the transaction fees, this orientation of the actual electricity consumption at the upper limit also means that, due to the periodic halvings, Bitcoin's electricity consumption will decrease significantly in the long term. At present, block rewards still account for about 80\,\% of the mining revenues. If the prices for electricity and Bitcoin and transaction fees remain unchanged, electricity consumption in 4,~8 and~12~years respectively would decrease by 40\,\%, 60\,\% and 70\,\% compared to the beginning of 2020 and settle at about 20\,\% of today's value, i.e., 20\,TWh/a and thus about ~4\,\% of today's German electricity consumption. However, economic conditions such as the prices for Bitcoin and transaction fees cannot be predicted reliably in practice. Nevertheless, it can be concluded that already in a few decades, and with limited Bitcoin prices, the value of the electrical energy consumed by Bitcoin will not be much higher than the cumulative transaction fees. This will in any case hold from 2140 onwards, but whether Bitcoin will be relevant for such a long period is difficult to assess after only 11 years. Due to the competition with other cryptocurrencies as well as technical developments, for example in terms of performance (reduction of the scarcity of transactions), it could be assumed that transaction fees will not increase significantly compared to today's level. This line of reasoning can be applied to many cryptocurrencies which, like Bitcoin, exhibit a periodic halving of block rewards.
\section*{Alternative consensus mechanisms}
\acresetall

In general, the power consumption of blockchains is made up of two components: The consensus mechanism, i.e. the process by which the nodes agree on which transactions are executed in which order, and the redundant execution of the transactions themselves, i.e. the verification of signatures and the adjustment of ``account balances'' in the local database of each node (``state transitions''). In Bitcoin, the \ac{pow} consensus mechanism as described above is responsible for the enormous power consumption; the cumulative effort for redundant transaction execution is negligible in comparison, even with the current size of the network of approximately 10\,000 nodes~\cite{deVries:2018:Bitcoin}. Since the development of Bitcoin, however, alternative consensus mechanisms have been developed, partly because of the high power consumption of \ac{pow}. For permissionless systems the most successful alternative so far is probably the \ac{pos} consensus mechanism. Here the voting weight is not linked to the resource computing power but to the resource capital (possessed units of the cryptocurrency, which is scarce and visibile and thus verifiable within the blockchain network). Often capital has to be ``frozen'' for a certain period of time in order to participate in voting and then serves as security to create an economic incentive for correct behaviour of the nodes. With \ac{pos} the competition for more and more computing power is thus eliminated. The energy requirement associated with the consensus mechanism is accordingly negligible compared to \ac{pow} and requires only a few operations on each node. Examples of current implementations of blockchains with \ac{pos} consensus mechanism are Algorand, Cardano, EOS and Tezos. In addition, Ethereum 2.0 (Serenity), the second largest cryptocurrency by market capitalization, will probably go into operation later this year. It is also based on \ac{pos} and is intended to replace or integrate the current Ethereum blockchain in the long term~\cite{ethhub}. Besides purely \ac{pos}-based blockchains there are also combinations of \ac{pos} and election-based protocols, such as the consensus mechanism Tendermint used for the Cosmos network. Such protocols also do not contain any particularly computationally or energy-intensive process.

There are often discussions among supporters of \ac{pow} and \ac{pos} whether \ac{pos} is as secure as \ac{pow} -- there are good arguments for both sides. For example, \ac{pow} can be expected to have long-term centralisation effects due to economies of scale in mining (procurement of hardware and electrical energy) and location-dependent economic framework conditions, which also reflects the current empirically observable centralisation of Bitcoin mining. This in turn may lead to security problems. With \ac{pos}, on the other hand, participation in the consensus mechanism corresponds to a return on the capital invested, so that the ratio of the capital of all participants in the consensus mechanism and thus the weight of their vote remains constant. On the other hand, with \ac{pow} it is also possible to participate in consensus without having to obtain resources from the network itself, which could make re-decentralisation considerably more difficult in an already highly centralised system. In addition, many \ac{pos} systems currently only allow one's own participation in the consensus mechanism if a certain minimum deposit is made, and for \ac{pow} blockchains there is also a simpler rule for deciding which blockchain is the ``valid'' in case of a conflict (fork).
The success of \ac{pos}-blockchains in the last years as well as research in this field~\cite{kiayias2017ouroboros} indicate that \ac{pos} can guarantee a level of security that is comparable to the level of security achieved in \ac{pow}.

Permissioned blockchains that are used in the context of consortia in the private and public sector all use voting-based consensus mechanisms, which can be regarded as partial crash-fault-tolerant simplifications (such as RAFT) or byzantine-fault-tolerant (such as PBFT or RBFT) successors of Paxos. As with \ac{pos}, their consensus-based energy consumption is thus also negligible. 

\section*{Redundant Operations}
Regardless of the type of consensus mechanisms, all blockchains are characterised by redundant data storage and operations. Accordingly, the cumulative computing effort and thus power consumption -- with the exception of hardware differences -- is proportional to the number of participants in the blockchain network.
For small blockchain networks, as they are typically used in consortia in a permissioned context, the redundancy results in a multiple power consumption compared to a central system. However, this does not necessarily mean that the use of a blockchain has to be a disadvantage from a sustainability perspective. The following rough estimate is intended to illustrate this: A small, private blockchain network, such as Hyperledger Fabric or Quorum, with 10 nodes, each with a mediocre hardware configuration (2~CPUs, 8GB~RAM) can easily handle 1000 simple transactions per second, thus each transaction consumes around 1\,J. On the other hand, based on the information provided in VISA's Sustainability Report 2017, it can be calculated that the energy consumption of the entire company (i.e. including heating of buildings, etc.) amounts to approximately 6\,000\,J per transaction, of which 3\,000\,J per transaction are incurred by data centres~\cite{VISA2017Sustainability}. A simple client-server system with a key-value store, such as LevelDB, can process several thousand simple transactions per second with the above hardware equipment, resulting in a power consumption in the order of 0.02\,J per transaction. Although the power consumption of a blockchain is generally much higher than that of a corresponding centralized solution (here by a factor of around 50) due to the redundancy (and to some extent also due to the consensus and generally the more extensive use of cryptographic methods), it may still only account for a very small part of the power consumption of the entire IT solution or the entire process, even if clients and backups are included. Particularly in scenarios in which processes can be further digitised with the help of energy-efficient variants of blockchain technology, it is therefore reasonable to assume that blockchain-based solutions can also save energy below the line.

In the best-known cryptocurrencies, such as Bitcoin and Ethereum, the corresponding blockchain networks already consist of many thousands of nodes, and with large-scale adoption, their number would be likely to rise even further in the future. Accordingly, the power consumption of these networks can be considered problematic because of the resulting high degree of redundancy. However, research also has promising solution concepts for this challenge: In principle, a reduced degree of redundancy, i.e. the recalculation of transactions only on subsets of all nodes, will also reduce the power consumption per transaction. This is the case, for example, with so-called second layer concepts such as Lightening or Raiden, but typically involves trade-offs in security, since security essentially stems from the high degree of redundancy. Similarly, in the Ethereum 2.0 blockchain mentioned above, the network will be divided into a total of 64 so-called shards, which will be periodically referenced by a main chain, the so-called ``Beacon Chain'', thus inheriting the security of the entire system with each such referencing. It will be some time before Ethereum 2.0 makes full use of these features and processes are distributed among the various shards, so it is difficult to quantify at this point how much the degree of redundancy will ultimately decrease and what impact this will have on security and functionality.

\section*{ZK-Rollups}
Particularly promising is the progress that has been made in recent years in connection with proofs of computational integrity using probabilistically checkable, succinct proofs, which are perhaps better known in the blockchain environment under the keyword ``Zero-Knowledge Proofs'' (ZKP). This makes it possible to show (probabilistically) with a usually very short and quickly verifiable proof that certain calculations have been performed correctly without having to specify all details of the calculation or to repeat the complete calculation. Initially, ZKPs were used by some cryptocurrencies such as Z-Cash to restore the confidentiality of transactions, which is practically non-existent with cryptocurrencies such as Bitcoin~\cite{zscashrepo}.


An essential characteristic of typical ZKPs is that the size of proofs and the computational complexity of verifying the proofs usually scale sublinearly (e.g. constant or polylogarithmic in the case of SNARKs or STARKs~\cite{gennaro2013quadratic, ben2019scalable}) with the size of the calculation to be verified. This makes it possible that a single party, e.g. a crypto exchange, bundles a large number (several thousand or even tens of thousand transactions) in so-called ``zk-rollups'' and sends only a short proof that all steps have been executed correctly (verification of signatures, correct updating of account balances, ...) as a transaction to the blockchain, where the proof is checked by the other nodes with little effort. In detail, the architectures can differ significantly, both in terms of the proof system used and the other data stored on the blockchain, which must be updated continuously (there is a trade-off here: less data on the blockchain means a higher dependency on the party creating the evidence and updating and storing the account balances of all accounts, but also higher scalability, as the blockchain then does not represent a bottleneck).

In contrast to existing second layer solutions, zk-rollups with complete on-chain data storage can provide the same security guarantees as the blockchain itself because the proof is still checked by all participants in the network and thus manipulations can be excluded with the same security as conventional transactions~\cite{optimisticvsrollup}. In addition, significant improvements can be achieved compared to the traditional processing of transactions, as the majority of the storage and computing capacity is based on digital signatures, the verification of which the operator of the zk-rollup compresses into a short proof. In existing prototypes, transaction rates of several hundred to several thousand transactions per second in Ethereum (conventionally approx. 10 transactions per second) have allegedly been already achieved with the help of zk-rollups~\cite{loopring2020rollup},~\cite{StarkDEX}. We want to estimate here with a (heavily simplified) example what implications this might have on the redundancy-related share of the power consumption of a large blockchain network.

On the basis of the zk-rollup of Loopring on the Ethereum blockchain, which is already in use, we can make a good estimate of the potential savings in the ideal case, i.e. with maximum utilisation of the zk-rollups and exclusive presence of transactions within the zk-rollup: For Loopring~3, the so-called ``gas'' costs, which are a measure of the storage and computing effort of a transaction in Ethereum and determine the pricing of the execution of transactions, are estimated to be around 365 at maximum utilisation of 2\,100 transactions per second~\cite{loopring2020rollup}. For comparison, a simple transaction in Ethereum requires at least 21\,000 gas, often significantly more. Accordingly, this already means a reduction of the redundant operations and thus their electricity demand per transaction by a factor of approx. 100 (an exact proportionality of gas costs and calculation effort is not given, but is nevertheless a reasonable approximation). On the other hand, the costs for the computationally complex proof generation, which only needs to be conducted by the operator of the zk-rollup, are estimated at 0.000042~USD per transaction~\cite{loopring2020rollup}. At 2\,100 transactions per second, this corresponds to an amount of approx.~5~USD per minute and thus, for example, the operation of AWS instances with approx. 96~vCPUs. For these, we can estimate a power consumption of some hundred watts~\cite{intel2020xeon, masanet2020recalibrating}, which does not result in more than 0.5\,J per transaction. This also corresponds to the order of magnitude from an alternative estimation, which, at least for ZKP-friendly hash functions, indicates an increased effort for the generation of evidence by a factor of around 100 compared to a simple calculation. On the other hand, own measurements with Ethereum-based blockchains without \ac{pow} consensus mechanism result in a power consumption of about 0.01\,J per transaction (tx) and node from CPU usage. In a network of approximately 10\,000 nodes, which is roughly the same size as the Bitcoin and Ethereum network today, we can therefore estimate a power consumption of approximately 100\,J/tx without zk-rollup. With zk-rollup, on the other hand, for the redundant operations, i.e. primarily proof verification, one obtains an amount of \mbox{(100/100 + 0.5)\,J/tx} and thus an energy saving of 98.5\,\%. For even larger networks, the energy saving in our example would further increase, whereas for smaller networks it would decrease and for networks with only very few nodes it would even lead to an increase in power consumption.

However, in a holistic view, the idle power consumption of nodes must also be taken into account. Depending on whether a separate computer is used or remains switched on exclusively for participation in the blockchain, or whether it is running anyway, or is located in a large cloud with a tendency to have a comparatively low idle consumption, as well as depending on the average load of the blockchain network, the idle consumption for large networks in non-\ac{pow} blockchains can significantly exceed the amount of energy consumed in connection with transactions or be negligible in comparison. Further improvements in energy proportionality could lead to idle consumption becoming less important in the future~\cite{shehabi2016united}; large data centres are generally more advanced in this respect than small ones~\cite{ruiu2017energy}. Furthermore, a scenario in which all transactions are processed within a zk-rollup is unrealistic. Nevertheless, it is conceivable that a large part of transactions can be processed within such zk-rollups in the future. Zk-rollups were primarily developed to solve scalability and performance problems of blockchains. Nevertheless, as just described, they can have the pleasant side effect of contributing to significant improvements in power consumption for large blockchain networks. However, these are only noticeable if no \ac{pow}-based consensus mechanism is already causing such high power consumption that improvements in redundant operations are not noticeable at all, and the idle consumption is negligible in absolute terms or in relation to the power consumption from CPU usage per transaction (this probably requires a considerable average load).
\section*{Conclusion}
\label{sec:conclusion}
The \ac{pow} blockchains and especially Bitcoin, which are currently used as the basis for many cryptocurrencies, have - considering their current technical performance - an enormous energy consumption. The total power consumption of all these \ac{pow} cryptocurrencies is still mainly caused by Bitcoin and amounts to between 20 -- 50\,\% of the German power consumption, with a best guess for Bitcoin at about 100\,TWh/a or 20\,\% of the German power consumption. The driving force behind the electricity consumption is the price of Bitcoin and not the number of transactions, and if the economic environment remains the same, the periodic halving of block rewards in many \ac{pow}-based cryptocurrencies would in the long term lead to a significant reduction in electricity consumption.

In addition, there are now established blockchains with alternative consensus mechanisms, above all \ac{pos} for public permissionless cryptocurrencies and the election-based consensus mechanisms of private permissioned blockchains. The latter usually include solutions that are used, for example, in companies or the public sector as a neutral platform across organisations. Due to the elimination of \ac{pow}, their power consumption is in each case orders of magnitude lower than that of Bitcoin and other \ac{pow}-based cryptocurrencies. However, mainly due to the redundant calculations characteristic of blockchain technology, their power consumption per transaction is roughly proportional to the number of participating nodes and thus still several times higher than that of central systems. Especially for large cryptocurrency networks, this can still mean high power consumption for non-\ac{pow} blockchains. Through technological further developments and modifications, with which the effort for redundant calculations and data storage can be reduced, and in particular zero knowledge proofs in zk-rollups, it is to be expected that in the future, the power consumption of large networks can be reduced further. However, the idle consumption must also always be taken into account when considering the whole network. 

\begin{figure}[!tbh]
    \centering
    \includegraphics[width=1.0\linewidth]{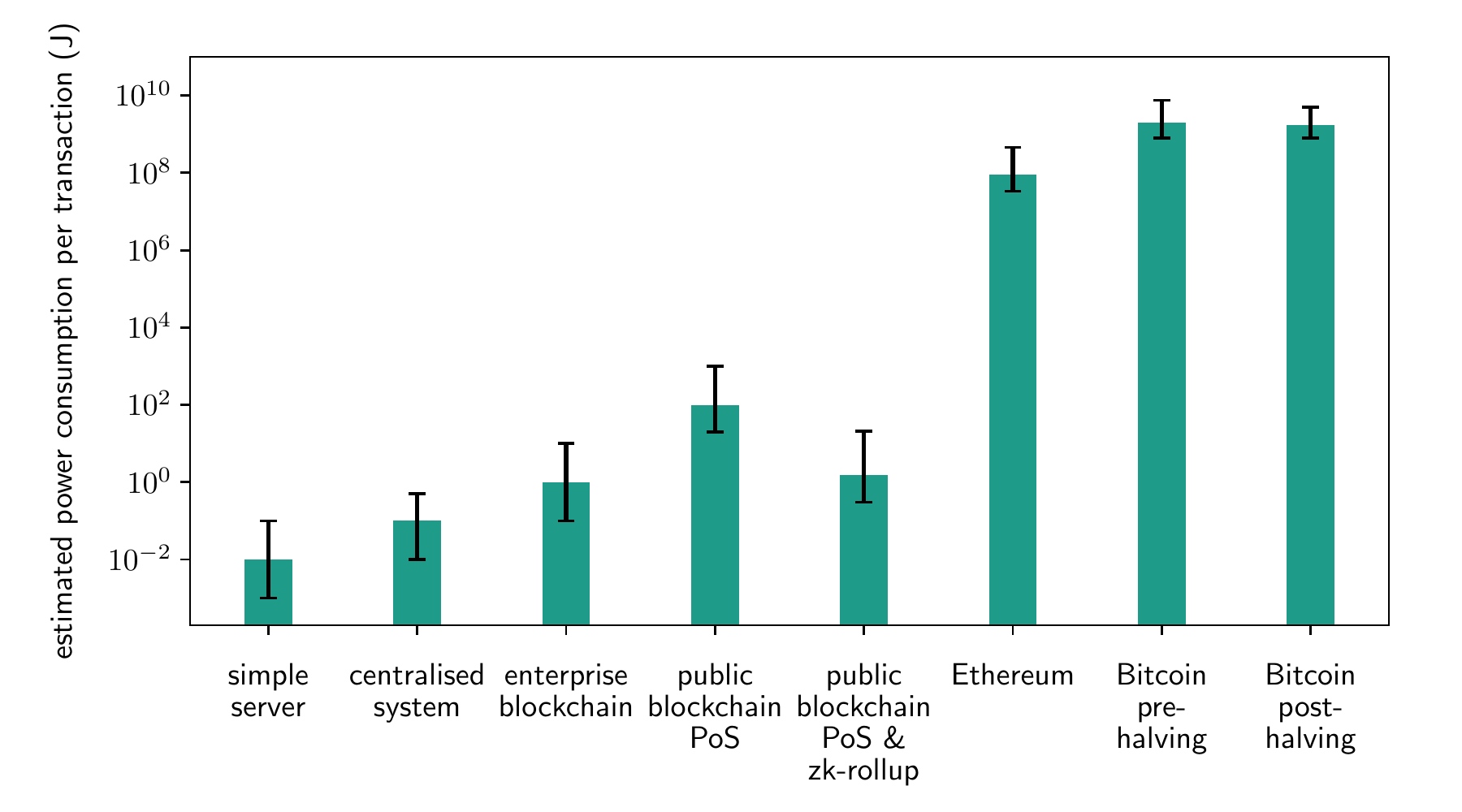}
    \caption{Rough comparison of the power consumption of different blockchain architectures per transaction. Note the logarithmic scale on the y-axis.}
    \label{fig:overall_comparison}
\end{figure}

Figure~\ref{fig:overall_comparison} summarises the findings of this paper by estimating typical power consumption of different blockchain technologies or improvements described in the previous sections. These are primarily for illustrative purposes and do not take the idle consumption into account. The figures given should therefore not yet be regarded as fixed and reliable, but only as an indication of the orders of magnitude. In particular, the error estimates are in most cases not generally reliable, as they correspond to empirical values from tests with few, different systems. However, the orders of magnitude are reasonable estimates based on the assumptions made in the respective chapters to the best of the authors' knowledge.
Figure~\ref{fig:overall_comparison} illustrates that the power consumption of blockchains already in use today is reduced by several orders of magnitude compared to the \ac{pow} blockchains of the ``first generation'' or could be with technology available today. 

Although the use of blockchain technology may not be the most energy-efficient solution from a purely technical point of view, what ultimately matters is the energy savings that can be achieved in return.
The potential in the cross-sectoral and cross-organizational use of
IT is particularly big~\cite{buhl2009green}. It is precisely here that many of the blockchain applications that go beyond cryptocurrencies and are currently being developed or tested in business and the public sector. For economic or political reasons, it is often not possible to implement a central, digital platform in these scenarios.  When using the described, energy-efficient blockchain solutions, it can be assumed that resources can be saved not only from a financial but also from an energy point of view when automating processes.

Here it is the task of business and information systems research to identify and quantify the existing potential for energy savings and climate protection and -- depending on the scenario with or without blockchain technology -- to exploit it with the most suitable technology. A prerequisite for this is that the consumption of resources (e.g. through a CO$_2$ tax) is priced in such a way that no ecologically harmful distortions occur and that the economic incentive mechanisms therefore promote the development of such solutions. In addition, it should always be kept in mind that a solution which initially appears expensive due to its high complexity and the still young technology, and which overall reduces the consumption of resources by using modern solutions such as blockchain, can also be a hedge against future price increases or fluctuations~\cite{buhl2011determining}.

In conclusion, we would therefore like to encourage further research both into technical improvements of blockchain technology, for example in terms of performance and energy efficiency, and into areas of application with particularly high potential for energy savings.

\begin{acknowledgements}
We would like to thank Peter Mertens for his valuable suggestions.
We also thank the Luxembourg National Research Fund (FNR) and PayPal for their support of the PEARL project ``P17/IS/13342933/PayPal-FNR/Chair in DFS/Gilbert Fridgen'', which made this article possible in the first place.
\end{acknowledgements}

\bibliographystyle{bise}

\bibliography{bib.bib}
\end{document}